\documentclass[journal]{IEEEtran}
\usepackage{mathrsfs}
\usepackage{url,times,amsmath,color,amssymb,graphicx,epsfig,psfig,cite,psfrag,subfigure,dsfont,booktabs}
\usepackage{algorithm}
\usepackage{algorithmic}
\usepackage{bm}

\begin{document}
\title{Energy Efficient User Association and Power Allocation in Millimeter Wave Based Ultra Dense Networks with Energy Harvesting Base Stations}

\author{Haijun Zhang,~\IEEEmembership{Member,~IEEE}, Site Huang, Chunxiao Jiang,~\IEEEmembership{Senior Member,~IEEE}, Keping Long,~\IEEEmembership{Senior Member,~IEEE}, Victor C. M. Leung,~\IEEEmembership{Fellow,~IEEE}, and H. Vincent
Poor,~\IEEEmembership{Fellow,~IEEE}

\thanks{Haijun Zhang and Keping Long are  with the Beijing Engineering and Technology Research Center for Convergence Networks and Ubiquitous Services, University of Science and Technology Beijing, Beijing, 100083, China. (e-mail: dr.haijun.zhang@ieee.org, longkeping@ustb.edu.cn).

Site Huang is with the College of Information Science and Technology, Beijing University of Chemical Technology, China (e-mail: huangsite378@gmail.com).

Chunxiao Jiang is with Tsinghu Space Center, Tsinghua University, Beijing 100084, P. R. China (e-mail: chx.jiang@gmail.com).

Victor C. M. Leung is with the Department of Electrical and Computer Engineering, The University of British Columbia, Vancouver, BC, V6T 1Z4, Canada (e-mail: vleung@ece.ubc.ca).

H. Vincent Poor is with the Department of Electrical Engineering, Princeton University, Princeton, NJ, USA (e-mail: poor@princeton.edu).

}} \maketitle
\begin{abstract}
Millimeter wave (mmWave) communication technologies have recently emerged as an attractive solution to meet the exponentially increasing demand on mobile data traffic. Moreover, ultra dense networks (UDNs) combined with mmWave technology are expected to increase both energy efficiency and spectral efficiency. In this paper, user association and power allocation in mmWave based UDNs is considered with attention to load balance constraints, energy harvesting by base stations, user quality of service requirements, energy efficiency, and cross-tier interference limits. The joint user association and power optimization problem is modeled as a mixed-integer programming problem, which is then transformed into a convex optimization problem by relaxing the user association indicator and solved by Lagrangian dual decomposition. An iterative gradient user association and power allocation algorithm is proposed and shown to converge rapidly to an optimal point. The complexity of the proposed algorithm is analyzed and the effectiveness of the proposed scheme compared with existing methods is verified by simulations.
\end{abstract}

\section{Introduction}

The proliferation of network devices and the growing demand for network services are contributing to a dramatic increase in overall network data traffic.
This problem is exacerbated in typical macrocell networks because of blind spots and shadowing. A promising solution is the deployment of ultra dense networks comprising flexibly deployed low-power small base stations (BSs), such as microcell BSs, picocell BSs and femtocell BSs \cite{Damnjanovic2011}.
In 5G \cite{MingXiaoTOM5G2016}, ultra dense networks that are deployed with low-cost and low-power small cells are expected to enhance the overall performance of the network in terms of energy efficiency and load balancing. The essence of ultra dense cell deployment is to shorten the physical distance between the transmitter and the receiver, so as to improve the performance of the system. Compared to traditional networks, ultra dense networks have the following advantages: (1) small cells can be deployed by users, which significantly reduces the cost of deployment; (2) ultra dense cells have a flexible configuration, and can reduce interference and improve energy efficiency through the setting of intelligent rules; and (3) ultra dense networks can completely solve problem of blind spots and achieve the goal of load-balancing.

However, in practice, due to the intensive characteristics of ultra dense deployment and the uncertainty of user deployment, radio resource allocation \cite{ZhouDong2016}, user association and interference mitigation become extremely challenging. Because of the high deployment density, cross-tier interference from small cell to macrocell \cite{Son2011}, local-tier interference from small cell to small cell, and the additive white Gaussian noise, cannot be ignored. These sources of interference can determine the association rule, as well as the power allocation policy between users and BSs. Under such circumstances, users are free to override the association's decisions to obtain a better payoff. On the other hand, the capacity of BSs has a limit, and thus we need to carefully consider these various elements in determining the association rules between users and BSs.

The millimeter wave (mmWave) frequency band is 30 \~{} 300 GHz, corresponding to wavelengths from 10 to 1 mm \cite{mmWave5G}. Due to its physical properties, mmWave can effectively solve many problems of high-speed broadband wireless access, and thus it has a broad application potential in short distance communication, such as in ultra dense small cell networks. The attenuation of mmWave reaches its maximum values in the 60 GHz, 120 GHz and 180 GHz bands \cite{MingXiaoTCOM2015}. This means that the interference levels for these attenuation bands are much lower compared to the 2-3 GHz bands. Moreover, due to the large 60 GHz bandwidth, mmWave systems can provide a relatively high data rates \cite{Athanasiou}. These characteristics of mmWave can play an indispensable role in enhancing spectral efficiency and energy efficiency of ultra dense networks.

Load-balancing is a main factor influencing the performance of BSs in heterogeneous ultra dense networks. Due to the cross-tier interference and the various capabilities of BSs, although users are uniformly associated with BSs, the uneven power allocation leads to differences in user experience \cite{ZhouMa2016} \cite{MingXiaoJsac2016}. Load awareness  based user association was proposed in \cite{YeUALB2013} and \cite{Andrews2014}. Traditionally, the signal-to-interference-plus-noise ratio (SINR) was utilized to determine whether a user should be associated with a given BS; however, this approach can lead to serious load-imbalance in the network. Load awareness transfers congested macrocells to lightly loaded small cells. Thus, although users are served by macrocells, they can make an association judgement and access a lightly loaded small cell.

Another important aspect of some ultra dense networks is the harvesting of energy from the radio-frequency (RF) environment to help power devices with limited energy resources \cite{Akbar2016}\cite{LiuKim2016}. Notably, wireless signals can transmit information and energy simultaneously, which means that transmitters can not only transmit information, but can also provide energy to charge the batteries of other devices. This method, known as simultaneous wireless information and power transfer (SWIPT) is a promising paradigm for ultra-dense networks.

\subsection{Related Work}

Many studies have addressed the problems of load-balancing and user association. In \cite{Shen14}, the authors proposed a proportional fairness scheme to solve the user association problem. The authors in \cite{Siddique2016} proposed a channel-access-aware association scheme in which the main idea is that each user estimates its channel access probability from different BSs to determine the best BS. Instead of calculating maximum SINR, this scheme gave preferential assignment to the low-power small cells. A particular power allocation scheme was proposed in \cite{Ghimire2013}, which presented a transmission coordination model called ON-OFF transmission coordination which can control the BS to transmit either the maximum transmit power or none. References \cite{ZhouLA2015}, \cite{Choi2010} and \cite{Fehske2010} all present load-aware schemes in user association for cellular networks, which provide guarantees on load-balancing and quality of service (QoS) requirements.

There have also been a number of studies of SWIPT, including [20]-[23]. In \cite{ChuZhu2016} and \cite{ZhangHuang2016}, the authors investigated secure communication in SWIPT with eavesdroppers and multiple energy harvesting receivers. In \cite{Lohani2016}, the authors studied the interference-aware resource allocation problem considering both time-switching and power-splitting approaches to SWIPT, in which interference signals were considered to improve the energy harvesting rate. Similarly, in \cite{LiuZhang2013}, the authors proposed different resource allocation schemes in SWIPT that guarantee a minimum the energy harvesting rate. In this paper, we will consider a form of SWIPT in which densely deployed base stations can harvest energy from the transmissions of other base stations to end users in the network.


There has also been some progress on user association in mmWave networks. In \cite{Distributedas}, the authors provided a distributed approach that solved the joint association and relaying problem in mmWave networks. Auction-based resource allocation in 60 GHz mmWave networks was presented in \cite{Athanasiou2013}. The authors of \cite{Semiari2015} proposed the use of matching theory for managing the spectrum resources of a heterogeneous small cell backhaul network. 

A goal of network design is to guarantee QoS in terms of user transmission rate. In \cite{Meng} and \cite{Zhuang}, the authors propose the use of historical information about QoS to assist users in selecting BSs with better long-term QoS instead of those with better immediate QoS. The literature has also addressed user association and energy efficiency in wireless networks. In \cite{Tabassum2014} and \cite{Blume2010}, the authors considered improving energy efficiency through base station sleeping technology. The authors in \cite{Beyranvand2015} proposed a distributed load-balancing algorithm for user association. In \cite{Dong2016}, the authors proposed a novel event data collection approach called reliability and multipath encounter routing (RMER), which can greatly reduce energy consumption in networks.
However, to the best of the authors' knowledge, user association and power allocation in mmWave based ultra dense networks, by jointly considering load-balancing constraints,  QoS requirements, energy efficiency, energy harvesting by base stations, and cross-tier interference limits, have not been studied in previous works.

\subsection{Main Contributions}

The main contributions in this paper can be summarized as follows:

\begin{itemize}
  \item Development of a novel energy efficient mmWave based ultra dense network optimization framework: This is a new approach to network optimization design based on the consideration of considering load-balancing constraints,  QoS requirements, energy efficiency, energy harvesting by base stations, and cross-tier interference limits in mmWave based ultra dense networks.
  \item Formulation of a user association and power allocation problem with multiple constraints: We formulate the load-aware energy efficient user association and power optimization problem in mmWave based ultra dense network as a mixed-integer programming problem. Cross-tier interference limits are used to protect the macrocells. A QoS requirement in terms of minimum achievable rate is provided to guarantee reliable transmissions.
  \item Proposal of load-aware energy efficient user association and power allocation algorithm: The formulated non-convex problem is transformed into a convex optimization problem by relaxing the user association indicator and solved by  Lagrangian dual decomposition. An iterative gradient user association and power allocation algorithm is proposed and is shown to converge to the optimal point quickly. The complexity of the proposed algorithm is analyzed and the effectiveness of the proposed scheme is verified by simulations comparing with existing method.
  \item Support of energy harvesting and quick convergence: Energy harvesting by base stations is considered in this paper. We also propose an iterative gradient method to update the transmit power variable and find an optimal association method for all users. We use the Newton-Raphson method to update the Lagrange multipliers and transmit power variable; it takes only a few iterations to converge towards an optimal point with guarantees on QoS and load-balancing. Theoretical analysis and simulation results verify the practicability and the feasibility of proposed method.
  \item Development of a load-aware scheme by using the log-utility function under power control: a load-aware scheme redistributes the traffic load to maximize the network utility and energy efficiency among BSs. Users tend to be attracted to macrocells with higher capacity for data transmission; the load-aware scheme reduces the priority of macrocells and redistributes the traffic selectively for all BSs so as to balance the load.
\end{itemize}

The rest of this paper is structured as follows. Section II introduces the system model, followed by a construction of the optimization problem formulation. Section III presents the Lagrangian dual decomposition formalism and an iterative gradient algorithm to find an optimal solution. Section IV includes an analysis of the performance of our proposed algorithms via simulations.  Section V concludes this paper.

\section{System Model and Problem Formulation}

\subsection{System Model}

As shown in Fig. 1, we consider a mmWave based ultra dense network, which consists of ultra dense small cells overlaid on one macrocell. We focus on the user association and transmit power allocation in the downlink of such a network. The set of all (macro and small) cells is denoted by $\mathbb{B} \in \left\{ {1,2,...,B} \right\}$, and the set of distributed users is denoted by $\mathbb{U} \in \left\{ {1,2,...,U} \right\}$. We assume that a user can be associate with only one BS, $\left\{ {{x_{ij}}} \right\}$ is used to indicate the association variable between user $i$ and BS $j$. If user $i$ is associated with BS $j$, ${x_{ij}} = 1$, otherwise ${x_{ij}} = 0$.

\begin{figure}[t]
        \centering
        \includegraphics*[width=80mm]{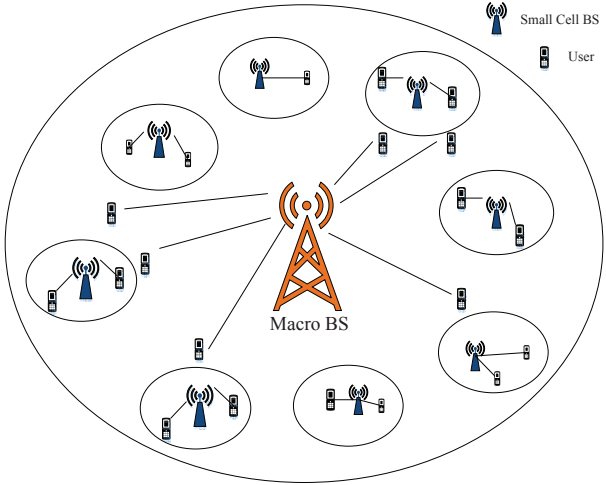}
        \caption{Architecture of a mmWave based ultra dense small cell networks with energy harvesting base stations.}
        \label{fig:1}
\end{figure}

Let ${g_{ij}}$ be the power gain from BS $j$ to user $i$. In this paper, we use the Friis transmission equation \cite{Mudumbai2015} to model the power gain:
\begin{equation}
{g_{ij}} = \frac{{g_{ij}^{{\text{Tx}}}g_{ij}^{{\text{Rx}}}{\varsigma ^2}}}{{16{\pi ^2}{{\left( {\frac{{{d_{ij}}}}{{{d_0}}}} \right)}^\eta }}}
\end{equation}
where ${g_{ij}^{{\text{Tx}}}}$ is the transmit antenna gain from BS $j$ to user $i$, ${g_{ij}^{{\text{Rx}}}}$ is the receive antenna gain from BS $j$ to user $i$, $\varsigma$ is the wavelength, ${{d_{ij}}}$ is the distance from BS $j$ to user $i$, ${{d_0}}$ is the \emph{far field reference distance}, and $\eta$ is the path-loss exponent ($\eta  \in \left[ {2,6} \right]$).

The SINR of user $i$ ($i \in {\mathbb{U}}$) receiving from the BS $j$ ($j \in {\mathbb{B}}$) can be written as
\begin{equation}
SIN{R_{ij}} = \frac{{{p_{ij}}{g_{ij}}}}{{\sum \limits_{k \in \mathbb{B},k \ne j} {{p_{kj}}{g_{kj}}}  + {\sigma ^2}}}
\end{equation}
where ${p_{ij}}$ is the transmit power from BS $j$ to user $i$, $W$ is the system bandwidth, and ${\sigma ^2}$ is the variance of additive white Gaussian noise (AWGN).
According to Shannon's capacity formula, the achievable rate for user $i$ from BS $j$ is given by
\begin{equation}
 c_{ij}  = \frac{W}{K_j}{\log _2}\left( {1 + SIN{R_{ij}}} \right),{\rm{  }}\forall {\rm{j}} \in {\mathbb{B}},
\end{equation}
where ${K_j}$ is the total number of users associated with BS $j$, and thus each user can receive $1/{K_j}$ of the total frequency band available.
As noted above, we assume that each user can be associated with only one BS. Because it is more difficult to implement multiple-BS association than single-BS association, multiple-BS association will increase the solution complexity and is not effective in practical systems. Thus, we can use ${x_{ij}}$ to be a binary variable to indicate whether user $i$ is associated with BS $j$.

\subsection{Problem Formulation}

Before modeling the problem, let us introduce the following constraints:

(1) User scheduling constraint: A user can be associated with only one BS at a time. Thus, we have\footnote{Unless otherwise denoted, summations over the BS index j extend over all of B, and summations over the user index i extend over all of U.}
\begin{equation}
\sum\limits_j {{x_{ij}} = 1,{\text{         }}\forall i \in {\mathbb{U}}}.
\end{equation}

(2) Total power constraint:
\begin{equation}
\sum\limits_i {{x_{ij}}{p_{ij}}}  \leqslant {p_{\max }},{\text{ }}\forall {j} \in {\mathbb{B}}
\end{equation}
where ${p_{\max }}$ denotes the maximum transmit power between user $i$ and BS $j$.

(3) QoS constraint:
\begin{equation}
\sum\limits_j {{x_{ij}}{c_{ij}} \geqslant {\text{ }}{R_t},{\text{    }}\forall i \in {\mathbb{U}}}
\end{equation}
where ${R_t}$ denotes the minimum transmit data rate for each user to maintain its own achievable rate.

(4) Cross-tier interference constraint:
\begin{equation}
\sum\limits_i {\sum\limits_{k \in \mathbb{B},k \ne j} {{x_{ij}}{p_{kj}}{g_{kj}}} }  \leqslant {I_j},{\text{  }}\forall {j} \in {\mathbb{B}}
\end{equation}
where ${I_j}$ denotes the maximum interference constraint. This constraint can be interpreted as an effective interference coordination mechanism. According to differences in traffic load conditions of small cells in ultra dense networks, the system will adjust the interference coordination mechanism dynamically to improve the spectral efficiency/energy efficiency.

(5) Constraint on the number of associated users:
\begin{equation}
\begin{gathered}
  \sum\limits_i {{x_{ij}}}  = {K_j},{\text{      }}\forall {j} \in {\mathbb{B}} \hfill \\
  where{\text{ }}0 \leqslant {K_j} \leqslant {N_U} \hfill. \\
\end{gathered}
\end{equation}

In order to maximize the network utility, we focus on the user association and resource allocation problem. Due to limited bandwidth, we assume that frequency resources are allocated for every associated user uniformly. In other words, we focus only on power allocation and user association to achieve load-balance and energy efficiency improvement.

During the process of user association with a BS, the issue of power consumption between the users and BSs needs to be considered. We also assume that each BS is equipped with an energy harvesting unit that can use received energy for replenishing a rechargeable battery.
With this assumption, the net power consumption of the network is given by
\begin{equation}
\begin{gathered}
  {U_P}\left( {X,P} \right){\text{ }} = {\text{ }}\sum\limits_j {{p_{cj}}}  + \sum\limits_i {\sum\limits_j {{x_{ij}}{p_{ij}}} }  \hfill \\
   \quad\quad\quad\quad\quad\quad - \psi\sum\limits_i {\sum\limits_j {\sum\limits_{m \in B} {  {x_{ij}}{p_{ij}}{{\left| {{g_{jm}}} \right|}^2}} } }  \hfill \\ 
\end{gathered} 
\end{equation}
where $X \triangleq {\left[ {{x_{ij}}} \right]_{{\mathbb{U}} \times {\mathbb{B}}}}$ is the user association matrix; $P \triangleq {\left[ {{p_{ij}}} \right]_{{\mathbb{U}} \times {\mathbb{B}}}}$ is the power allocation matrix; $\sum\limits_j {{p_{cj}}}$ is the circuit power consumed by the BSs; the term $\sum\limits_i {\sum\limits_j {{x_{ij}}{p_{ij}}} }$ is the total transmit power consumed by the BSs, which will be particularly affected by the user association problem; and $\psi\sum\limits_{m \in B} {  {x_{ij}}{p_{ij}}{{\left| {{g_{jm}}} \right|}^2}} $ is the energy harvested by all BSs, where $\psi$ is the coefficient of energy harvesting efficiency of the BSs.

So given the power consumption function, we can formulate the optimization problem of interest as follows:
\begin{equation}
\mathop {\max }\limits_{X,P} {\text{  }}\sum\limits_i {\sum\limits_j {{x_{ij}}\frac{{\log \left( {\frac{W{\log _2}\left( {1 + SIN{R_{ij}}} \right)}{{\sum\nolimits_{k \in \mathbb{U}} {{x_{kj}}} }}} \right)}}{{{U_P}\left( {X,P} \right)}}} }
\end{equation}
\begin{equation}
\begin{aligned}
& \text{s.t.}
& & {\begin{gathered}C1:{U_P}\left( {X,P} \right){\text{ }} = {\text{ }}\sum\limits_j {{p_{cj}}}  + \sum\limits_i {\sum\limits_j {{x_{ij}}{p_{ij}}} }  \hfill \\
   \quad\quad\quad\quad\quad\quad\quad\quad-\psi \sum\limits_i {\sum\limits_j {\sum\limits_{m \in B} {  {x_{ij}}{p_{ij}}{{\left| {{g_{jm}}} \right|}^2}} } }  \hfill \\ 
\end{gathered}}  \\
&&&{C2:\sum\limits_j {{x_{ij}} = 1,{\text{         }}\forall i \in {\mathbb{U}}}} \\
&&& {C3:\sum\limits_i {{x_{ij}}}  = {K_j},{\text{      }}\forall {j} \in {\mathbb{B}}} \\
&&& {C4:\sum\limits_i {{x_{ij}}{p_{ij}}}  \leqslant {p_{\max }},{\text{ }}\forall {j} \in {\mathbb{B}}}\\
&&& {C5:\sum\limits_j {{x_{ij}}{c_{ij}} \geqslant {\text{ }}{R_t},{\text{    }}\forall i \in {\mathbb{U}}}}\\
&&& {C6:{x_{ij}} \geqslant 0,{\text{ }}0 \leqslant {K_j} \leqslant {N_U}{\text{    }}\forall i \in {\mathbb{U}},{\text{ and }}\forall {j} \in {\mathbb{B}}}\\
&&& {C7:\sum\limits_i {\sum\limits_{k \in \mathbb{B},k \ne j} {{x_{ij}}{p_{kj}}{g_{kj}}} }  \leqslant {I_j},{\text{  }}\forall {j} \in{\mathbb{B}}}\\
\end{aligned}
\end{equation}
where $C1$ is the total power consumption according to (9); $C2$ guarantees that each user can be associated with only one BS; $C3$ is the constraint that there are ${K_j}$ users being served by BS $j$; $C4$ is the
maximum transmit power limit of user $i$ from BS $j$; $C5$ sets the QoS requirement ${R_t}$ for user $i$ to ensure its achievable rate; $C6$ specifies the ranges of ${x_{ij}}$ and ${K_j}$; and $C7$ is the cross-tier interference constraint.

\section{Lagrangian Dual Decomposition}

In this section, we consider the solution to the optimization problem (10)-(11). This problem is a mixed-integer optimization problem, which has very high complexity due to the lack of convexity of the objective and the binary nature of the variables in $X.$. To address the latter problem, we relax these variables, and replace (10)-(11) with the following:
\begin{equation}
\mathop {\max }\limits_{X,P} {\text{  }}\sum\limits_i {\sum\limits_j {{x_{ij}}\frac{{\log \left( {\frac{W{\log _2}\left( {1 + SIN{R_{ij}}} \right)}{{\sum\nolimits_{k \in \mathbb{U}} {{x_{kj}}} }}} \right)}}{{{U_P}\left( {X,P} \right)}}} }
\end{equation}
\begin{equation}
\begin{aligned}
& \text{s.t.}
& & {\begin{gathered}C1:{U_P}\left( {X,P} \right){\text{ }} = {\text{ }}\sum\limits_j {{p_{cj}}}  + \sum\limits_i {\sum\limits_j {{x_{ij}}{p_{ij}}} }  \hfill \\
   \quad\quad\quad\quad\quad\quad\quad\quad- \psi\sum\limits_i {\sum\limits_j {\sum\limits_{m \in B} { {x_{ij}}{p_{ij}}{{\left| {{g_{jm}}} \right|}^2}} } }  \hfill \\ 
\end{gathered}}  \\
&&& {C2:\sum\limits_j {{x_{ij}} = 1,{\text{         }}\forall i \in {\mathbb{U}}}} \\
&&& {C3:\sum\limits_i {{x_{ij}}}  = {K_j},{\text{      }}\forall {j} \in {\mathbb{B}}} \\
&&& {C4:\sum\limits_i {{x_{ij}}{p_{ij}}}  \leqslant {p_{\max }},{\text{ }}\forall {j} \in {\mathbb{B}}}\\
&&& {C5:\sum\limits_j {{x_{ij}}{c_{ij}} \geqslant {\text{ }}{R_t},{\text{    }}\forall i \in {\mathbb{U}}}}\\
&&& {C6:{x_{ij}} \geqslant 0,{\text{ }}0 \leqslant {K_j} \leqslant {N_U}{\text{    }}\forall i \in {\mathbb{U}},{\text{ and }}\forall {j} \in {\mathbb{B}}}\\
&&& {C7:\sum\limits_i {\sum\limits_{k \in \mathbb{B},k \ne j} {{x_{ij}}{p_{kj}}{g_{kj}}} }  \leqslant {I_j},{\text{  }}\forall {j} \in{\mathbb{B}}}\\
&&& {C8:{\text{0}} \leqslant {x_{ij}} \leqslant 1,{\text{        }}\forall i \in {\mathbb{U}},{\text{ and }}\forall {j} \in {\mathbb{B}}}\\.
\end{aligned}
\end{equation}
We use the Lagrangian dual decomposition method to solve the relaxed problem (12)-(13). The corresponding Lagrangian function is given by
\begin{equation}
\begin{gathered}
  L(\left\{ {{x_{ij}}} \right\},\left\{ {{p_{ij}}} \right\},\mu ,\lambda ,\nu ,\tau ){\text{  }} \hfill \\
   = \sum\limits_{i \in {\mathbb{U}}} {\sum\limits_{j \in {\mathbb{B}}} {{x_{ij}}\frac{{\log \left( {\frac{W{\log _2}\left( {1 + SIN{R_{ij}}} \right)}{{\sum\nolimits_{k \in \mathbb{U}} {{x_{kj}}} }}} \right)}}{{{U_P}\left( {X,P} \right)}}} } {\text{  +  }}\sum\limits_{j \in {\mathbb{B}}} {{\mu _j}\left( {{K_j} - \sum\limits_{i \in {\mathbb{U}}} {{x_{ij}}} } \right)} {\text{ }} \hfill \\
  {\text{ +  }}\sum\limits_{j \in {\mathbb{B}}} {{\lambda _j}\left( {{p_{\max }} - \sum\limits_{i \in {\mathbb{U}}} {{x_{ij}}{p_{ij}}} } \right)} {\text{  + }}\sum\limits_{i \in {\mathbb{U}}} {{\nu _i}\left( {\sum\limits_{j \in {\mathbb{B}}} {{x_{ij}}{c_{ij}}}  - {R_t}} \right)}  \hfill \\
   + \sum\limits_{{\text{j}} \in {\mathbb{B}}} {{\tau _j}\left( {{I_j} - \sum\limits_{i \in {\mathbb{U}}} {\sum\limits_{k \in \mathbb{B},k \ne j} {{x_{ij}}{p_{kj}}{g_{kj}}} } } \right)}  \hfill \\
\end{gathered}
\end{equation}
where $\mu  = {\left[ {{\mu _1},{\mu _2},...,{\mu _B}} \right]^T}$, $\lambda  = {\left[ {{\lambda _1},{\lambda _2},...,{\lambda _B}} \right]^T}$, $\nu  = {\left[ {{\nu _1},{\nu _2},...,{\nu _U}} \right]^T}$ and $\tau  = {\left[ {{\tau _1},{\tau _2},...,{\tau _B}} \right]^T}$ are the Lagrange multipliers used to relax the coupled constraint. Thus, the Lagrangian dual function can be written as
\begin{equation}
{\text{D}}\left( {\mu{\text{,}}\lambda{\text{,}}\nu{\text{,}}\tau } \right) = \mathop {\max }\limits_{X,P} {\text{ }}L(\left\{ {{x_{ij}}} \right\},\left\{ {{p_{ij}}} \right\},\mu ,\lambda ,\nu ,\tau ),
\end{equation}
and so the Lagrangian problem of (15) can be written as
\begin{equation}
\mathop {\min }\limits_{\mu,\lambda,\nu } {\text{   D}}\left( {\mu{\text{,}}\lambda{\text{,}}\nu{\text{,}}\tau } \right).
\end{equation}

\subsection{Dual Decomposition}

We divide the original problem into two independent subproblems through the Lagrangian dual method, so that we can solve this optimization problem by solving the two subproblems.

The problem (12) subject to the constraints (13) can be rewritten as
\begin{equation}
\mathop {\max }\limits_{X,P} {\text{  }}\left\{ \begin{gathered}
  \sum\limits_i {\sum\limits_j {{x_{ij}}\frac{{\log \left[ {W{{\log }_2}\left( {1 + SIN{R_{ij}}} \right)} \right]}}{{{U_P}\left( {X,P} \right)}}} } {\text{ }} \hfill \\
  \quad\quad\quad - {\text{ }}\sum\limits_i {{K_j}\frac{{\log \left( {{K_j}} \right)}}{{{U_P}\left( {X,P} \right)}}}  \hfill \\ 
\end{gathered}  \right\}
\end{equation}

where we let the power consumption $C1$ be evenly distributed on both fractional sides. The dual problem using the Lagrangian dual decomposition method is given by
\begin{equation}
\mathop {\min }\limits_{\mu,\lambda,\nu,\tau} {\text{   D}}\left( {\mu{\text{,}}\lambda{\text{,}}\nu{\text{,}}\tau } \right){\text{  =  }}{f_{X,P}}\left( {\mu ,\lambda ,\nu ,\tau } \right){\text{ }} + {\text{ }}{g_{K,P}}\left( {\mu ,\lambda ,\nu ,\tau } \right)
\end{equation}
where
\begin{equation}
\begin{gathered}
  {f_{X,P}}\left( {\mu ,\lambda ,\nu ,\tau } \right){\text{  =  }} \hfill \\
  \left\{ \begin{gathered}
\mathop {\max }\limits_{x,p} {\text{   }}\left\{ \begin{gathered}
  \sum\limits_i {\sum\limits_j {{x_{ij}}\frac{{\log \left[ {W{{\log }_2}\left( {1 + SIN{R_{ij}}} \right)} \right]}}{{{U_P}\left( {X,P} \right)}}} }  \hfill \\
   - \sum\limits_i {\sum\limits_j {{x_{ij}}\left( {{\mu _j}{\text{   +   }}{\lambda _j}{p_{ij}} - {\nu _i}{c_{ij}}} \right)} }  \hfill \\
   - \sum\limits_i {\sum\limits_j {{x_{ij}}{\tau _j}\sum\limits_{k \in {\mathbb{B}},k \ne j} {{p_{kj}}{g_{kj}}} } }  \hfill \\ 
\end{gathered}  \right\}\hfill \\ 
  s.t.{\text{      }}\sum\limits_j {{x_{ij}}}  = 1 \hfill \\
  {\text{           0 }} \leqslant {\text{ }}{x_{ij}}{\text{ }} \leqslant {\text{1}} \hfill \\
 {\begin{gathered}{U_P}\left( {X,P} \right){\text{ }} = {\text{ }}\sum\limits_j {{p_{cj}}}  + \sum\limits_i {\sum\limits_j {{x_{ij}}{p_{ij}}} }  \hfill \\
   \quad\quad\quad\quad\quad\quad- \psi\sum\limits_i {\sum\limits_j {\sum\limits_{m \in B} {  {x_{ij}}{p_{ij}}{{\left| {{g_{jm}}} \right|}^2}} } }  \hfill \\ 
\end{gathered}}  \\
\end{gathered}  \right. \hfill \\
\end{gathered}
\end{equation}
\begin{equation}
\begin{gathered}
  {g_{K,P}}\left( {\mu ,\lambda ,\nu ,\tau } \right){\text{  =  }} \hfill \\
  \left\{ \begin{gathered}
\mathop {\max }\limits_{K,p} {\text{   }}\left\{ \begin{gathered}
  \sum\limits_j {\left( {{\mu _j}{K_j}{\text{ }} + {\text{ }}{\lambda _j}{p_{\max }}{\text{ }} - {\text{ }}\sum\limits_i {{\nu _i}{R_t}}  + {\tau _j}{I_j}} \right)}  \hfill \\
   - \sum\limits_i {{K_j}\frac{{\log \left( {{K_j}} \right)}}{{{U_P}\left( {X,P} \right)}}}  \hfill \\ 
\end{gathered}  \right\}\hfill \\ 
  s.t.{\text{      }}{K_j}{\text{ }} \leqslant {\text{ }}{N_U} \hfill \\
 {\begin{gathered}{U_P}\left( {X,P} \right){\text{ }} = {\text{ }}\sum\limits_j {{p_{cj}}}  + \sum\limits_i {\sum\limits_j {{x_{ij}}{p_{ij}}} }  \hfill \\
   \quad\quad\quad\quad\quad\quad - \psi\sum\limits_i {\sum\limits_j {\sum\limits_{m \in B} {  {x_{ij}}{p_{ij}}{{\left| {{g_{jm}}} \right|}^2}} } }  \hfill \\ 
\end{gathered}}  \\
\end{gathered}  \right. \hfill \\
\end{gathered}
\end{equation}
The optimization problems (17) and (18) are the
problems with respect to $\left( {X,P} \right)$ and $\left( {\mu ,\lambda ,\nu } \right)$ respectively. Thus, the original problem in (12) under condition (13) is transformed into subproblems (19) and (20).

The partial derivative of the objective of (19) can be expressed as
\begin{equation}
\begin{gathered}
  \frac{{\partial {f_{X,P}}\left( {\mu ,\lambda ,\nu ,\tau } \right)}}{{\partial {x_{ij}}}}{\text{ }}
  {\text{ =  }}\frac{{\log \left[ W{\log _2}\left( {1 + SIN{R_{ij}}} \right) \right]}}{{{U_P}\left( {X,P} \right)}}{\text{ }} \hfill \\ \quad \quad - {\text{ }}{\mu _j}{\text{ }} - {\text{ }}{\lambda _j}{p_{ij}} + {\nu _i}{c_{ij}}
- {\tau _j}\sum\limits_{k \in \mathbb{B},k \ne j} {{p_{kj}}{g_{kj}}}  \hfill \\
\end{gathered}
\end{equation}
In order to achieve the maximum of (17), the maximizer $\left\{ {{x_{ij}}} \right\}$ of the subproblem (19) is defined as
\begin{equation}
{x_{ij}} = \left\{ \begin{gathered}
  1,{\text{   }}if{\text{  }}j = {j^*} \hfill \\
  0,{\text{  }}if{\text{  }}j \ne {j^*} \hfill \\
\end{gathered}  \right.
\end{equation}
where
\begin{equation}
\begin{gathered}
  {j^*}{\text{  =  }} \hfill \\
  \mathop {\arg \max }\limits_j \left( \begin{gathered}
  \frac{{\log \left[ W{\log _2}\left( {1 + SIN{R_{ij}}} \right) \right]}}{{{U_P}\left( {X,P} \right)}}{\text{ }}  - {\text{ }}{\lambda _j}\left( t \right){p_{ij}}\left( t \right)  \hfill \\
  - {\text{ }}{\mu _j}\left( t \right) + {\nu _i}\left( t \right){c_{ij}}- {\tau _j}\left( t \right)\sum\limits_{k \in \mathbb{B},k \ne j} {{p_{kj}}\left( t \right){g_{kj}}}  \hfill \\
\end{gathered}  \right) \hfill \\
\end{gathered}
\end{equation}
At the $k$th inner iteration, (23) can be considered to be a judgement criterion for users to determine the best service or the highest network utility of the BS.

Similarly, we can obtain ${K_j}$ from the partial derivative of (20):
\begin{equation}
{K_j}\left( {t + 1} \right){\text{  =  }}{e^{\left[ {{\mu _j}\left( {\text{t}} \right) \cdot {U_P}^{\left( t \right)}\left( {X,P} \right){\text{ }} - {\text{ 1}}} \right]}}
\end{equation}
where ${K_j}$ can be considered to be an optimum association scheme. That is, it can be used by BS $j$ to choose the specific number of users with which to associate.

We use the subgradient method to update the Lagrange multipliers as follows \cite{Haijun2015}:
\begin{equation}
{\mu _j}\left( {t + 1} \right){\text{  =  }}{\mu _j}\left( {\text{t}} \right){\text{ }} - {\text{ }}{\delta _1}\left( t \right) \cdot \left( {{K_j}\left( t \right){\text{ }} - {\text{ }}\sum\limits_i {{x_{ij}}\left( t \right)} } \right)
\end{equation}
\begin{equation}
{\lambda _j}\left( {t \!\!+\!\! 1} \right){\text{  \!\!= \!\!  }}{\lambda _j}\left( t \right){\text{ }} \!\!-\!\! {\text{ }}{\delta _2}\left( t \right) \cdot \left( {{p_{\max }}{\text{ }} \!\!-\!\! {\text{ }}\sum\limits_i {{x_{ij}}\left( t \right){p_{ij}}\left( t \right)} } \right)
\end{equation}
\begin{equation}
{\nu _i}\left( {t + 1} \right){\text{  =   }}{\nu _i}\left( t \right){\text{ }} - {\text{ }}{\delta _3}\left( t \right) \cdot \left( {\sum\limits_j {{x_{ij}}\left( t \right){c_{ij}}} {\text{ }} - {\text{ }}{R_t}} \right)
\end{equation}
\begin{equation}
{\tau _j}\left( {t \!\!+\!\! 1} \right){\text{  \!\!=\!\!   }}{\tau _j}\left( t \right){\text{ }} \!\!\!\!\!-\!\!\!\! {\text{ }}{\delta _4}\left( t \right) \cdot \left( {{I_j} \!\!-\!\! \!\sum\limits_i {\sum\limits_{k \in \mathbb{B},k \ne j} {{x_{ij}}\left( t \right){p_{kj}}\left( t \right){g_{kj}}} } } \!\!\right)
\end{equation}
where ${\delta _1}\left( t \right)$, ${\delta _2}\left( t \right)$, ${\delta _3}\left( t \right)$ and ${\delta _4}\left( t \right)$ are step sizes. By updating the Lagrange multipliers ${\mu _j}\left( {\text{t}} \right)$, ${\lambda _j}\left( t \right)$, ${\nu _i}\left( t \right)$ and ${\tau _j}\left( t \right)$ via (25)-(28), the dual problem will achieve the global optimum when the multipliers converge.

In fact, we can use the Lagrange multiplier ${\mu _j}$ (the price of the BSs for users) to choose the best service through \emph{the law of supply and demand}: ${K_j}$ represents the best standard; if the service demand $\sum\limits_i {{x_{ij}}\left( t \right)} $ exceeds the supply ${K_j}$, that will lead to higher price ${\mu _j}$ to balance the market supply and demand. The users compare the obtained payoff from different associated BSs to determine whether the BS $j$ is suitable to associate with through the judgement scheme (25). In other cases, the price will be affected by the loads of the BSs. If a BS is overloaded, then it will have to increase its price.

\subsection{Energy Efficiency and Power Allocation}

In this subsection, we use the Newton-Raphson method to solve the power optimization problem.

We can rewrite the Lagrangian function (14) to get the power optimization function as:
\begin{equation}
\begin{aligned}
\begin{gathered}
  f({p_{ij}}) = \sum\limits_{i \in {\mathbb{U}}} {\sum\limits_{j \in {\mathbb{B}}} {{x_{ij}}\frac{{{{\log }_e}\left( {\frac{{{{\log }_2}\left( {1 + SINR} \right)}}{{{K_j}}}} \right)}}{{{p_{cj}} + \left( {1 - \sum\limits_{m \in B} {\psi  \cdot {x_{ij}}{{\left| {{g_{jm}}} \right|}^2}} } \right) \cdot {p_{ij}}}}} } {\text{ }} \hfill \\
  {\text{            \qquad  \quad \quad +  }}\sum\limits_{j \in {\mathbb{B}}} {{\lambda _j}\left( {{p_{\max }} - \sum\limits_{i \in {\mathbb{U}}} {{x_{ij}}{p_{ij}}} } \right)} {\text{ }}. 
\end{gathered}
\end{aligned}
\end{equation}
We assume user $i$ associates with BS $j$. Therefore, under the joint association, ${x_{ij}}$ can be regarded as a constant value of 1.
Then, the first-order partial derivative and the second derivative of the power function (29) are given by (30) and (31) at the top of next page.
\newcounter{TempEqCnt}
     \setcounter{TempEqCnt}{\value{equation}}
    \begin{figure*}[ht]
 \begin{equation}\label{powerSolution}
\begin{array}{*{20}{l}}
\frac{{\partial f}}{{\partial {p_{ij}}}} = \sum\limits_{i \in \mathbb{U}} {\sum\limits_{j \in \mathbb{B}} {\frac{{SINR\left( {{p_{cj}} + \left( {1 - \sum\limits_{m \in B} {\psi {x_{ij}}{{\left| {{g_{jm}}} \right|}^2}} } \right){p_{ij}}} \right)}}{{\ln 2{{\left( {{p_{cj}} + \left( {1 - \sum\limits_{m \in B} {\psi {x_{ij}}{{\left| {{g_{jm}}} \right|}^2}} } \right){p_{ij}}} \right)}^2}}}} }  \cdot \frac{1}{{{p_{ij}}\left( {1 + SINR} \right){{\log }_2}\left( {1 + SINR} \right)}} - \sum\limits_{i \in \mathbb{U}} {\sum\limits_{j \in \mathbb{B}} {\frac{{\left( {1 - \sum\limits_{m \in B} {\psi {x_{ij}}{{\left| {{g_{jm}}} \right|}^2}} } \right){{\log }_e}\left( {\frac{{{{\log }_2}\left( {1 + SINR} \right)}}{{{K_j}}}} \right)}}{{{{\left( {{p_{cj}} + \left( {1 - \sum\limits_{m \in B} {\psi {x_{ij}}{{\left| {{g_{jm}}} \right|}^2}} } \right){p_{ij}}} \right)}^2}}}} }  \hfill \\  \quad \quad \quad - \sum\limits_{j \in \mathbb{B}} {{\lambda _j}}
\end{array}
\end{equation}\hrulefill
\end{figure*}


\newcounter{TempEqCnt2}
     \setcounter{TempEqCnt2}{\value{equation}}
     \begin{figure*}[ht]
 \begin{equation}\label{powerSolution}
\begin{array}{*{20}{l}}
\begin{gathered}
\begin{gathered}
  \frac{{{\partial ^2}f}}{{\partial {p_{ij}}^2}} = \sum\limits_{i \in {\mathbb{U}}} {\sum\limits_{j \in {\mathbb{B}}} {\frac{{SINR}}{{\left( {{p_{cj}} + \left( {1 - \sum\limits_{m \in B} {\psi  \cdot {x_{ij}}{{\left| {{g_{jm}}} \right|}^2}} } \right){p_{ij}}} \right){p_{ij}}\ln 2}}} }  \cdot \frac{{ - SINR \cdot \left( {1 + \ln 2 \cdot {{\log }_2}\left( {1 + SINR} \right)} \right)}}{{\ln 2 \cdot {p_{ij}}{{\left( {1 + SINR} \right)}^2}{{\left( {{{\log }_2}\left( {1 + SINR} \right)} \right)}^2}}} \hfill \\
   - \sum\limits_{i \in {\mathbb{U}}} {\sum\limits_{j \in {\mathbb{B}}} {\frac{{SINR}}{{\left( {{p_{cj}} + \left( {1 - \sum\limits_{m \in B} {\psi  \cdot {x_{ij}}{{\left| {{g_{jm}}} \right|}^2}} } \right){p_{ij}}} \right){p_{ij}}\ln 2}}} }  \cdot \frac{{SINR \cdot \left( {1 - \sum\limits_{m \in B} {\psi  \cdot {x_{ij}}{{\left| {{g_{jm}}} \right|}^2}} } \right)}}{{\ln 2 \cdot {p_{ij}}\left( {1 + SINR} \right){{\log }_2}\left( {1 + SINR} \right)}} \hfill \\
   - \sum\limits_{i \in {\mathbb{U}}} {\sum\limits_{j \in {\mathbb{B}}} {\frac{{2\left( {1 - \sum\limits_{m \in B} {\psi  \cdot {x_{ij}}{{\left| {{g_{jm}}} \right|}^2}} } \right)}}{{{{\left( {{p_{cj}} + \left( {1 - \sum\limits_{m \in B} {\psi  \cdot {x_{ij}}{{\left| {{g_{jm}}} \right|}^2}} } \right) \cdot {p_{ij}}} \right)}^3}}}} }  \cdot \frac{{SINR \cdot \left( {{p_{cj}} + \left( {1 - \sum\limits_{m \in B} {\psi  \cdot {x_{ij}}{{\left| {{g_{jm}}} \right|}^2}} } \right) \cdot {p_{ij}}} \right)}}{{\ln 2 \cdot {p_{ij}}\left( {1 + SINR} \right){{\log }_2}\left( {1 + SINR} \right)}} \hfill \\
   - {\text{ }}\sum\limits_{i \in {\mathbb{U}}} {\sum\limits_{j \in{\mathbb{B}}} {\frac{{2\left( {1 - \sum\limits_{m \in B} {\psi  \cdot {x_{ij}}{{\left| {{g_{jm}}} \right|}^2}} } \right)}}{{{{\left( {{p_{cj}} + \left( {1 - \sum\limits_{m \in B} {\psi  \cdot {x_{ij}}{{\left| {{g_{jm}}} \right|}^2}} } \right) \cdot {p_{ij}}} \right)}^3}}}} }  \cdot \left( {1 - \sum\limits_{m \in B} {\psi  \cdot {x_{ij}}{{\left| {{g_{jm}}} \right|}^2}} } \right) \cdot {\log _e}\left( {\frac{{{{\log }_2}\left( {1 + SINR} \right)}}{{{K_j}}}} \right) \hfill \\
\end{gathered}
\end{gathered}
\end{array}
\end{equation}\hrulefill
\end{figure*}

Due to computational complexity of introducing the Hessian matrix, we focus only on the two-dimensional case.

In this case, the variable ${p_{ij}}$ is updated as follows:
\begin{equation}
{p_{ij}}\left( {t + 1} \right) = {p_{ij}}\left( t \right) + {\delta _4}\left( t \right)\Delta {p_{ij}}.
\end{equation}
where ${\delta _4}\left( t \right)$ is the step size, and $\Delta p_{ij }$ is given by
\begin{equation}
\Delta {p_{ij}} = {{\left| {\frac{{\partial f}}{{\partial {p_{ij}}}}} \right|} \mathord{\left/
 {\vphantom {{\left| {\frac{{\partial f}}{{\partial {p_{ij}}}}} \right|} {\left| {\frac{{{\partial ^2}f}}{{\partial {p_{ij}}^2}}} \right|}}} \right.
 \kern-\nulldelimiterspace} {\left| {\frac{{{\partial ^2}f}}{{\partial {p_{ij}}^2}}} \right|}}
\end{equation}
From the QoS constraint (6), we can obtain the minimum transmit power as
\begin{equation}
\begin{gathered}
  {{\overset{\lower0.5em\hbox{$\smash{\scriptscriptstyle\smile}$}}{p} }_{ij}} = \frac{{{I_{ij}}}}{{{g_{ij}}}}\left( {{2^{Rt}} - 1} \right) \hfill \\
  {\rm If} {\text{ }}{p_{ij}} < {{\overset{\lower0.5em\hbox{$\smash{\scriptscriptstyle\smile}$}}{p} }_{ij}}, {\rm then}{\text{ }}{p_{ij}} = {{\overset{\lower0.5em\hbox{$\smash{\scriptscriptstyle\smile}$}}{p} }_{ij}}. \hfill \\.
\end{gathered}
\end{equation}
By updating the transmit power variable, the value of the net power consumption is also changed. This is how we control the power variable to influence the user association scheme. Once we update the power control formula to reach convergence, the communication system will achieve a near-optimal load balancing situation. At this moment, we define the user association scheme as load-aware association, which can transfer the traffic from a congested macrocell to relatively low-load or lightly loaded small cells. Thus, during the next trading time, the almost overloaded BSs will change their prices to reduce the number of associated users. And users will re-select objects based on the prices of the BSs. This load-aware association scheme is superior to SINR-based association. In the SINR-based algorithm, the BS overload problem is not considered.

\subsection{Iterative Gradient Algorithm}

In this subsection, we propose an iterative gradient algorithm to find the optimal user association solution.

{\renewcommand\baselinestretch{1.0}
\begin{algorithm}[!t]\vspace{-1pt}
\small
\caption{Iterative Gradient Algorithm}
\begin{algorithmic}[1]
\STATE  Initialization: ${I_{\max }}$ and ${p_{\max }}$, Let ${p_{ij}} = {p_0}$
\STATE  Initialization: Set the Lagrange multipliers ${\mu _j}$, ${\lambda _j}$ and ${\nu _i}$ to zero
\STATE  Set $i = 0$
\REPEAT
\STATE  User association
\FOR    {$i = 1$ to $\mathbb{U}$}
\FOR    {$j = 1$ to $\mathbb{B}$}
\STATE  (1)	Calculate the power consumption ${U_P}\left( {P,X} \right)$ according to (9);
\STATE  (2)	Calculate ${j^*}$ according to (23);
\STATE  (3)	Use the ${j^*}$ to update ${x_{ij}}$ according to (22);
\STATE  (4)	Update ${\mu _j}$ according to (25);
\STATE  (5)	Update ${\lambda _j}$ according to (26);
\STATE  (6)	Update ${\nu _i}$ according to (27);
\STATE  (7)	Update ${\tau _j}$ according to (28);
\ENDFOR
\ENDFOR
\STATE  Power allocation
\FOR    {$i = 1$ to $\mathbb{U}$}
\FOR    {$j = 1$ to $\mathbb{B}$}
\STATE  (1)	Calculate $\frac{{\partial f}}{{\partial {p_{ij}}}}$ and $\frac{{{\partial ^2}f}}{{\partial {p_{ij}}^2}}$;
\STATE  (2)	Update $\Delta {p_{ij}}$ according to (32), find the optimal step size ${\delta _4}\left( t \right)$ according to \cite{Convex};
\STATE  (3)	${p_{ij}}\left( {t + 1} \right) = {p_{ij}}\left( t \right) + {\delta _4}\left( t \right)\Delta {p_{ij}}$;
\IF {${\text{ }}{p_{ij}} < {\overset{\lower0.5em\hbox{$\smash{\scriptscriptstyle\smile}$}}{p} _{ij}}$ do}
\STATE  ${p_{ij}} = {\overset{\lower0.5em\hbox{$\smash{\scriptscriptstyle\smile}$}}{p} _{ij}}$;
\ENDIF
\IF {$\sum\limits_i {{x_{ij}}{p_{ij}}}  < {p_{\max }},{\text{ }}\forall {\text{j}} \in {\mathbb{B}}$ do}
\STATE  ${p_{ij}} = {p_{\max }}$;
\ENDIF
\STATE  Update ${p_{ij}}$;
\ENDFOR
\ENDFOR
\UNTIL {Convergence or $i = {I_{\max }}$};

\end{algorithmic}
\end{algorithm}
\par}
This algorithm is shown as Algorithm 1, which solves the association problem (12) under the constraints in (13). In updating the variable ${p_{ij}}$, we use Newton's method which has a relatively fast convergence rate. Algorithm 1, by firstly fixing the power variable ${p_{ij}}$ to update the Lagrange multipliers and then updating the power variable ${p_{ij}}$, is feasible for practical use.

\subsection{Complexity Analysis}

In this subsection, we analyze the complexity of the proposed gradient algorithm. In Algorithm 1, the association scheme needs $o\left( {B \times U} \right)$ operations to establish the associated relationship between users and BSs at each iteration, and power allocation also need $o\left( {B \times U} \right)$ operations. We suppose that the optimal solution to the problem (12) requires $I$ iterations to converge. All users at each iteration will receive price lists of all BSs and the communication service package (23) from the base stations. Then they will choose one of the BSs to associate with, which can guarantee their minimum rates. From the BS perspective, the BS will update the price list ${\mu _j}$ according to the standard of achievable rates for users and load-balancing. Each of the Lagrange multipliers ${\mu _j}$, ${\lambda _j}$ and ${\tau _j}$ require $o\left( B \right)$ operations, and ${\nu _i}$ requires $o\left( U \right)$ operations. So the complexity of Algorithm 1 is $o\left( {I\left( {2BU + 3B + U} \right)} \right)$.

\section{Simulation results and discussion}

In this section, we compare our proposed gradient algorithm with the MAX-SINR association algorithm by analyzing their performance via simulation. We consider an ultra dense network and all users are uniformly distributed within one macro cell. We set the deployed densities of small cells and users as $\left\{ {{\lambda _B},{\lambda _U}} \right\} = \left\{ {1500,6000} \right\}$ per macrocell. The radius of the macrocell is 100 m. The additive white Gaussian noise power is set as ${\sigma ^2} = kTB =  - 134{\text{dBm}}$. The maximum transmitting power of the macrocell is set at 9.5 dBm, and the maximum transmitting powers of the small cells are all set at 4.7 dBm. We set $\varsigma  = 5{\text{ mm}}$, ${d_0} = 1{\text{ m}}$ and $W = 1200{\text{ MHz}}$. Moreover, for all $i \in \mathbb{U},j \in \mathbb{B}$, we set ${p_{ij}} = {p_0} = 0.1{\text{mW}}$ and $g_{ij}^{{\text{Tx}}} = g_{ij}^{{\text{Rx}}} = 1$. Finally, we set the energy harvester efficiency as $\psi  = 0.8$.

\begin{figure}[h]
        \centering
        \includegraphics*[width=80mm]{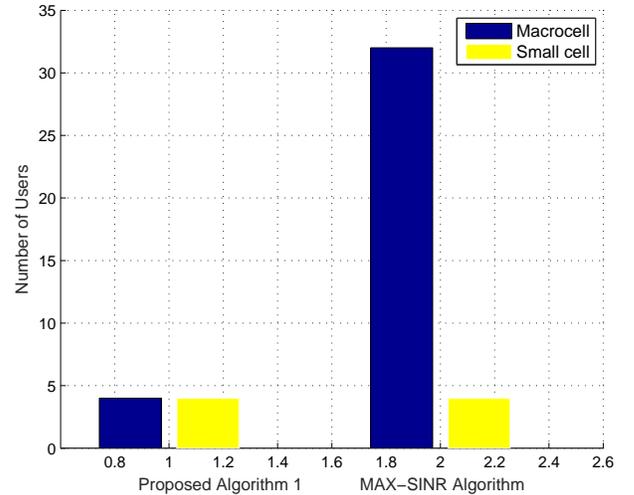}
        \caption{The numbers of users associated with the macrocell and with each small cell under different association algorithms.}
        \label{fig:2}
\end{figure}
Fig. 2 compares the number of users served by the macrocell and each small cell under different association schemes. The MAX-SINR association scheme shows that many users will be associated with the macrocell and it leads to a seriously unbalanced load, since the macrocell has a higher transmit power. By contrast, our proposed gradient algorithm promotes a load-balancing and energy-efficient association scheme. The proportions of users associated with the macrocell and each small cell are equal under our proposed algorithm. This reduces the macrocell load pressure and transfers congested users to a lightly loaded small cell in order to improve the overall network's energy efficiency.

\begin{figure}[h]
        \centering
        \includegraphics*[width=80mm]{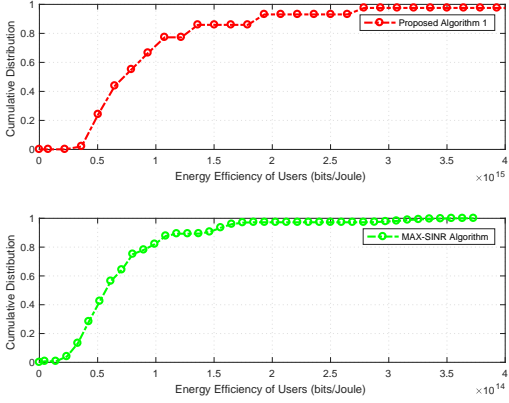}
        \caption{Cumulative distribution function of energy efficiency of users}
        \label{fig:3}
\end{figure}
\begin{figure}[h]
        \centering
        \includegraphics*[width=80mm]{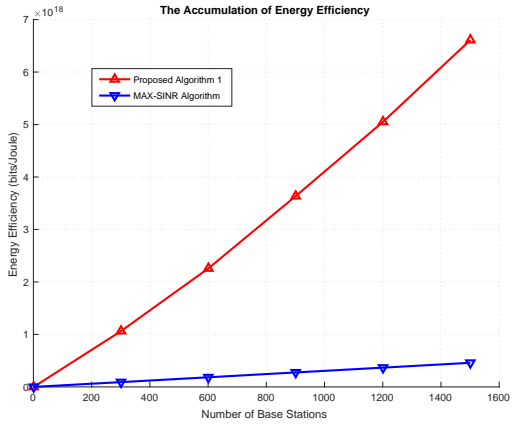}
        \caption{The accumulation of energy efficiency}
        \label{fig:4}
\end{figure}
Fig. 3 and Fig. 4 consider performance from the user perspective and small cell perspective, respectively, to compare the energy efficiency of the two different algorithms. From the Fig. 3, under our proposed algorithm, the range of energy efficiency of users is mostly distributed between $0.5 \times {10^{15}}$ and $3 \times {10^{15}}$ bits/Joule. However, the range of energy efficiency of users under the MAX-SINR algorithm is from $0.3 \times {10^{14}}$ to $1.8 \times {10^{14}}$ bits/Joule. From this numerical comparison, it is concluded that the energy efficiency of users resulting from our proposed algorithm is 10 times that of the MAX-SINR algorithm. Fig. 4 shows that the sum of energy efficiency achieved by our gradient algorithm is also almost 10 times that of the MAX-SINR association algorithm. Because the load capacity of each base station is limited, our algorithm seeks to maximize the energy efficiency \cite{YeUAIM2016} through coordinating the relationship between the system load and user's minimum service rate \cite{HanMIAO}. However, the MAX-SINR algorithm aims to maximize the user rate regardless of the load-balancing and energy efficiency, and for this reason, it will not usually attain a high rate of energy efficiency. So we can conclude that our proposed gradient method improves significantly in energy efficiency compared to the MAX-SINR method.

\begin{figure}[h]
        \centering
        \includegraphics*[width=80mm]{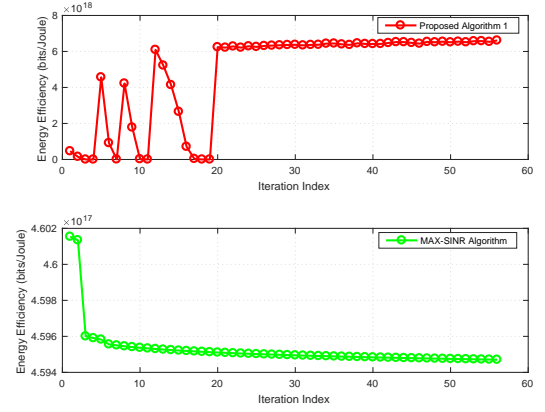}
        \caption{Convergence behavior of energy efficiency.}
        \label{fig:5}
\end{figure}
Fig. 5 shows the convergence behavior of energy efficiency of the two association schemes. We use Newton's method, which has a satisfactory convergence rate, to solve the problem (12) under constraints (13).  From Fig. 5, we can draw an obvious conclusion that the proposed gradient based user association method requires approximately 20 iterations to reach the optimal point, but the MAX-SINR method requires approximately 35 iterations to find its optimal solution. Moreover, from Fig. 5, we see that for the gradient method the energy efficiency, or utility of power, is higher compared with the MAX-SINR method at each iteration. Moreover, as the number of iterations increases, the energy efficiency of the MAX-SINR algorithm is gradually reduced.
So our proposed algorithm can not only achieve a near-optimal point, but it can also provide a load-balancing property in order to maximize the network utility under the power control constraint and QoS requirements. As the load aware association scheme addresses the network's energy efficiency, according to \cite{Liu2015}, to maximize the network's energy efficiency, it tends to lead to proportional fairness. This means that reducing the macrocell power consumption facilitates load-balancing and higher energy efficiency.

\begin{figure}[h]
        \centering
        \includegraphics*[width=80mm]{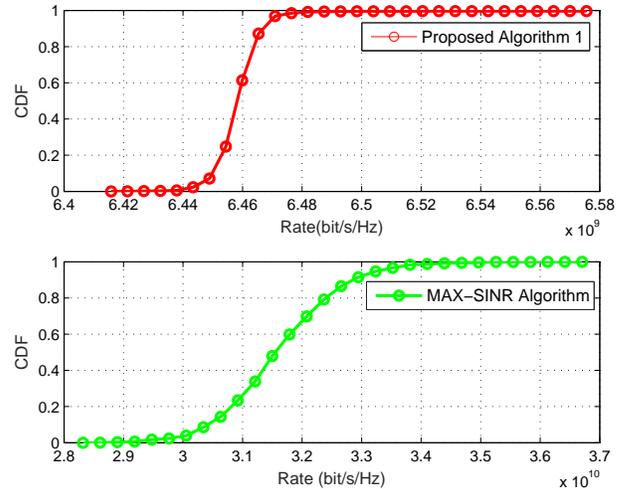}
        \caption{CDFs of user rates under different user association schemes.}
        \label{fig:6}
\end{figure}
Fig. 6 shows the cumulative distribution functions (CDFs) of user long-term rates of the two association schemes. We set the QoS constraint ${R_t}$ to 1 bps/Hz for both the proposed algorithm and MAX-SINR algorithm.
User rates for MAX-SINR association range from $2.9 \times {10^{10}}$ to $3.4 \times {10^{10}}$ bits/s/Hz, but most of the achievable rates for users range from $6.44 \times {10^9}$ to $6.48 \times {10^9}$ bits/s/Hz in our proposed algorithm. The MAX-SINR algorithm is effective in enhancing the user rate. But the gap between user rates of the two algorithms is not that dramatic. And the rate that we get from our proposed algorithm satisfies the minimal user rate. Moreover, our proposed algorithm can provide users with a relatively constant value, which means that the pricing rules of all base stations (including the macro base station) are similar and constant. All users in this network can find the optimal associated base station that satisfies their minimal user rate and offers a reasonable price. Thus, this approach solves the problems of the limitation of macrocell traffic load and the existence of blind spots in the macrocell. The macrocell cannot provide an ideal achievable rate for all users. So an ultra dense network together with our proposed algorithm solves these problems. 

\begin{figure}[h]
        \centering
        \includegraphics*[width=80mm]{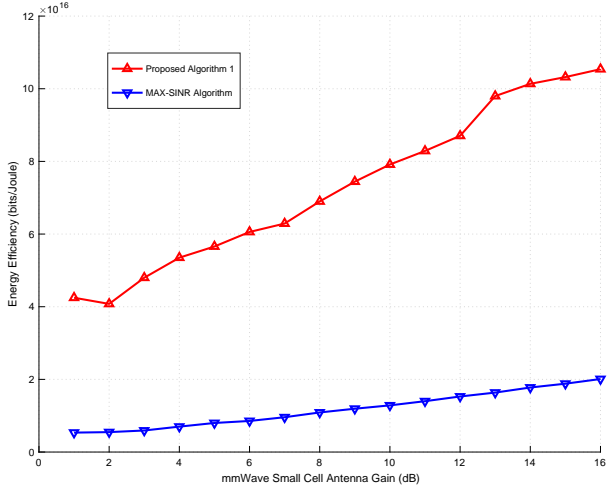}
        \caption{Energy efficiency versus the mmWave small cell antenna gain.}
        \label{fig:7}
\end{figure}

We now consider the blockage effect in the mmWave channel model. According to \cite{Ghosh2014}, we have correspondingly corrected the path loss as $PL\left( d \right) = 20{\log _{10}}\left( {\frac{{4\pi {d_0}}}{\varsigma }} \right) + 10\eta {\log _{10}}\left( {\frac{d}{{{d_0}}}} \right) + {\sigma ^2}$. A simple yet accurate channel model is applied for simplifying blockage modeling according to \cite{Bai2015}, where a user within a certain distance from a BS is considered to be in line-of-sight (LOS) contact and beyond that distance is assumed to be in non-line-of-sight (NLOS) contact. The LOS and NLOS path loss exponents for BS-to-user are set as 2 and 3.4, respectively, and the LOS and NLOS path loss exponents for BS-to-BS are set as  2 and 3.5, respectively. The LOS and NLOS shadowing factors for BS-to-user are set as 5.9 and 7.6, respectively, and the LOS and NLOS shadowing factors for BS-to-BS are set as 6.5 and 7.9, respectively.

Fig. 7 shows the antenna (omni-directional) gain's effects on the energy efficiency when considering blockage in the mmWave channel model. As can be seen in Fig. 7, our proposed algorithm has higher energy efficiency than the MAX-SINR algorithm in this case as well provides more energy efficiency than MAX-SINR algorithm.

\begin{figure}[h]
        \centering
        \includegraphics*[width=80mm]{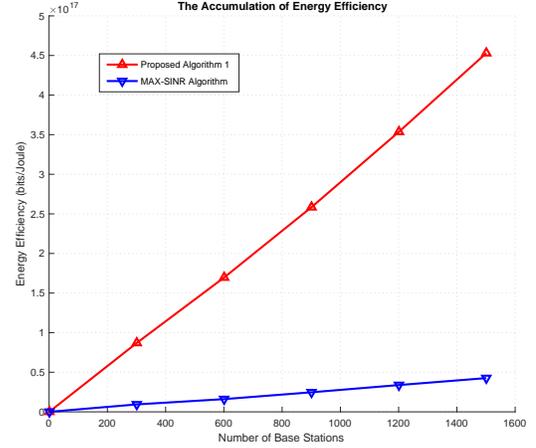}
        \caption{The accumulation of energy efficiency considering the blockage effect in the mmWave channel for the two algorithms.}
        \label{fig:8}
\end{figure}
Fig. 8 shows the accumulation of energy efficiency from the small cell perspective when the antenna gain is set as 13 dBi. As the number of base stations increases, the accumulation of energy efficiency also increases. As can be seen from the figure,  when the number of base stations is 1500, our proposed algorithm's energy efficiency is about 9 times that of the traditional MAX-SINR algorithm. Therefore, even with blockage taken into consideration, our proposed algorithm still has significant advantage in achieving energy efficiency.

\begin{figure}[h]
        \centering
        \includegraphics*[width=80mm]{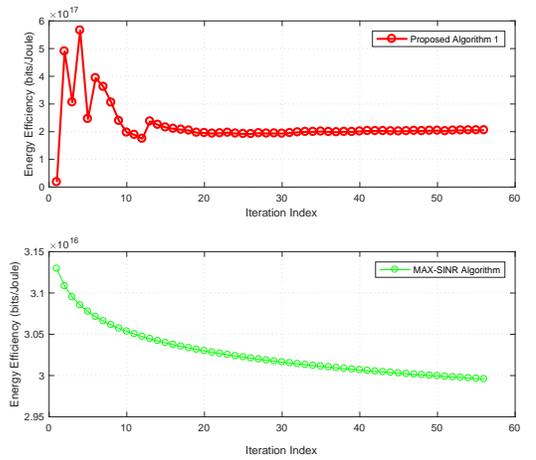}
        \caption{Convergence behavior of energy efficiency with blockage effect in the mmWave channel model for the two algorithms.}
        \label{fig:9}
\end{figure}
Fig. 9 shows the convergence behavior of energy efficiency of the two user association schemes when considering the blockage effect. It shows that our proposed algorithm still converges rapidly. Our proposed algorithm requires approximately 20 iterations to converge, while the energy efficiency curve of the MAX-SINR algorithm seems to not convergence due to its neglect of load balancing.


\section{Conclusion}

In this paper, we have considered a mmWave based ultra dense network, which also combines energy harvesting at base stations. We have proposed an effective interference coordination mechanism to cognitively limit the interference between the BSs and users in ultra dense networks. Moreover, we have modeled a network utility optimal function under constraints on power and QoS. We have also proposed a gradient association technique to solve this optimization problem, which has a satisfactory convergence rate and can find a near-optimal solution. By using Lagrangian dual decomposition, the dual optimization problem can be decoupled into two sub-problems, which can be solved separately.

The simulation results show that our algorithm outperforms the MAX-SINR algorithm. The numerical results demonstrate that load-balancing association and price control user association significantly improve the network utility and energy efficiency, and it also provides a satisfactory user experience in pricing rules and user rate under the minimal user rate constraint. 

\section*{Acknowledgment}
This work was supported by the National Natural Science Foundation of China (Grant 61471025), the Chinese Association of Science and Technology Young Elite Scientist Sponsorship Program, the Chinese Fundamental Research Funds for the Central Universities, and the U. S. National Science Foundation under Grant ECCS-1343210.

\begin{IEEEbiography}[{\includegraphics[width=1in,height=1.25in,clip,keepaspectratio]{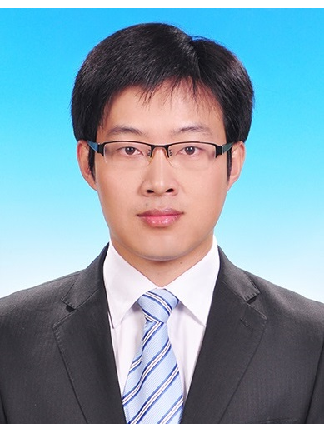}}]{Haijun Zhang}(M'13)
is currently a Full Professor in University of Science and Technology Beijing, China. He was a Postdoctoral Research Fellow in Department of Electrical and Computer Engineering, the University of British Columbia (UBC), Vancouver Campus, Canada. He received his Ph.D. degree in Beijing University of Posts Telecommunications (BUPT). From 2011 to 2012, he visited Centre for Telecommunications Research, King's College London, London, UK, as a Visiting Research Associate. Dr. Zhang has published more than 80 papers and authored 2 books. He serves as Editor of IEEE 5G Tech Focus, Journal of Network and Computer Applications, Wireless Networks, Telecommunication Systems, and KSII Transactions on Internet and Information Systems, and serves/served as Leading Guest Editor for IEEE Communications Magazine, IEEE Transactions on Emerging Topics in Computing and ACM/Springer Mobile Networks \& Applications. He serves/served as General Co-Chair of 5GWN'17 and 6th International Conference on Game Theory for Networks (GameNets'16), Track Chair of 15th IEEE International Conference on Scalable Computing and Communications (ScalCom2015), Symposium Chair of the GameNets'14, and Co-Chair of Workshop on 5G Ultra Dense Networks in ICC 2017.
\end{IEEEbiography}

\begin{IEEEbiography}[{\includegraphics[width=1in,height=1.25in,clip,keepaspectratio]{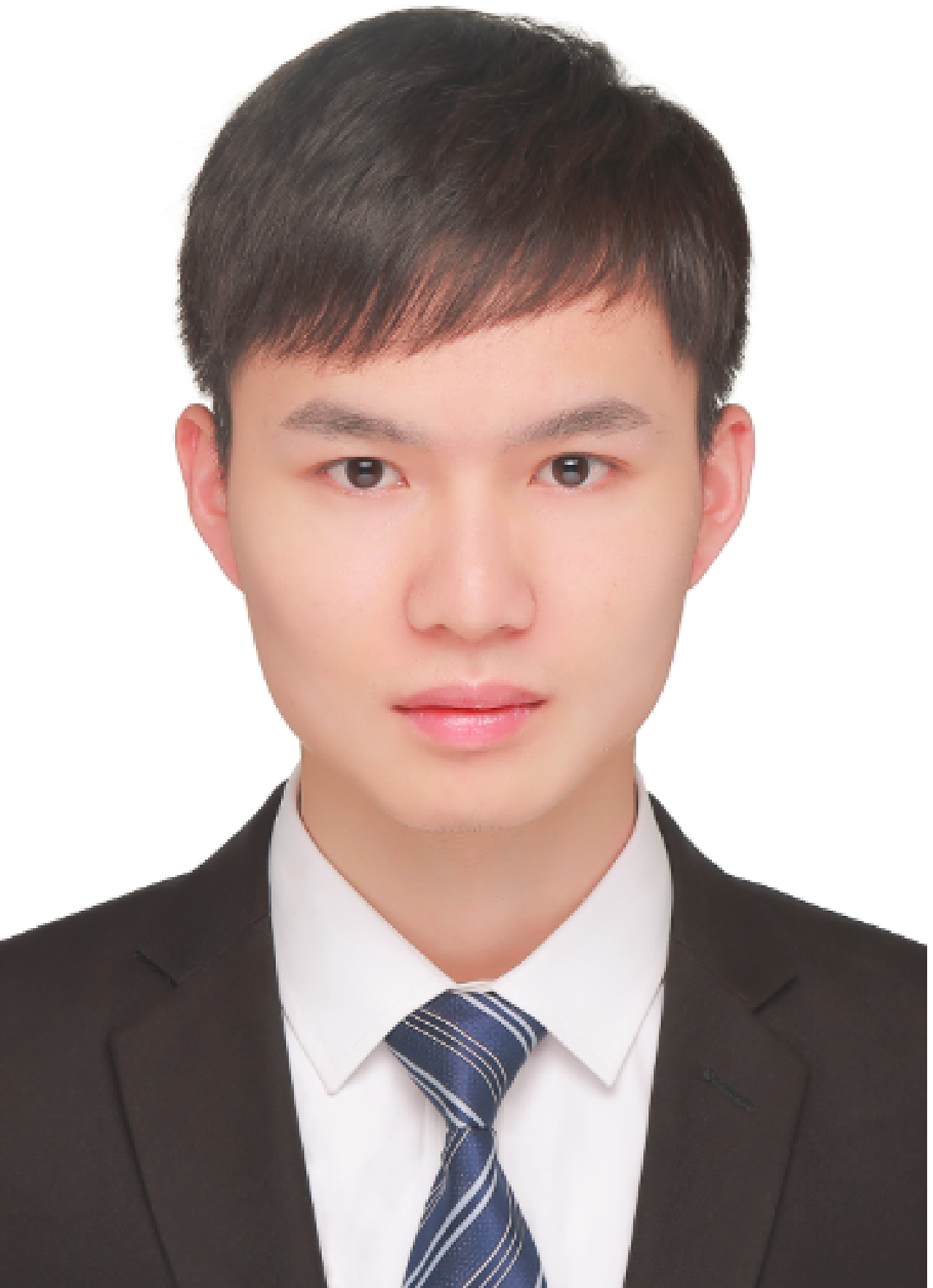}}]{Site Huang}
is currently pursuing the BS degree in communication engineering from Beijing University of Chemical Technology, Beijing, China. He did research on 5G and ultra dense networks as a research assistant in the University of Science and Technology Beijing, China. His research interests include resource allocation, power control, and energy efficiency in wireless communication.
\end{IEEEbiography}

\begin{IEEEbiography}[{\includegraphics[width=1in,height=1.25in,clip,keepaspectratio]{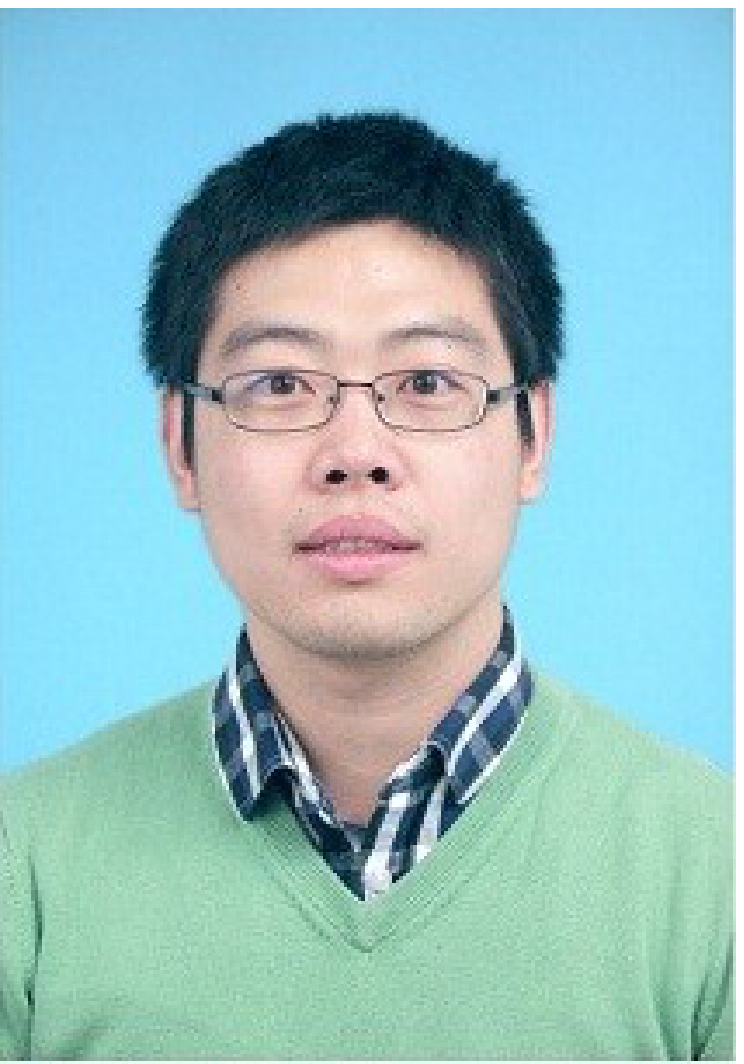}}]{Chunxiao Jiang}(S'09-M'13-SM'15) received the B.S. degree in information engineering from Beihang University, Beijing in 2008 and the Ph.D. degree in electronic engineering from Tsinghua University, Beijing in 2013, both with the highest honors. Currently, he is an assistant research fellow in Tsinghua Space Center, Tsinghua University. His research interests include applications of game theory, optimization, and statistical theories to communication, networking, signal processing, and resource allocation problems, in particular space information networks, heterogeneous networks, social networks, and big data privacy. He was the recipient of the Best Paper Award from IEEE GLOBECOM in 2013, the Best Student Paper Award from IEEE GlobalSIP in 2015, the Distinguished Dissertation Award from CAAI (Chinese Association for Artificial Intelligence) in 2014 and the Tsinghua Outstanding Postdoc Fellow Award (only ten winners each year) in 2015. He is a senior member of IEEE.
\end{IEEEbiography}

\begin{IEEEbiography}[{\includegraphics[width=1in,height=1.25in,clip,keepaspectratio]{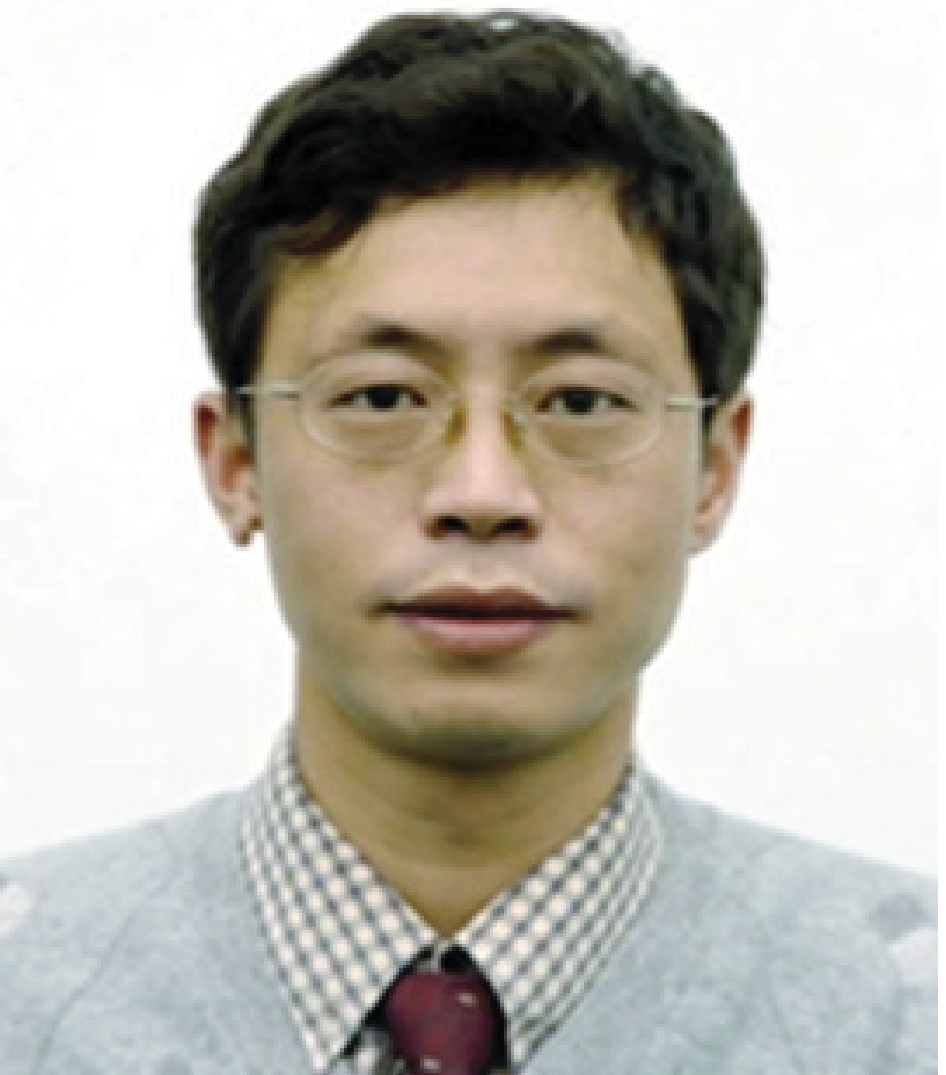}}]{Keping Long} (SM'06) received the M.S. and Ph.D. degrees from the University of Electronic Science and Technology of China, Chengdu, in 1995 and 1998, respectively.

From September 1998 to August 2000, he was a Postdoctoral Research Fellow at the National Laboratory of Switching Technology and Telecommunication Networks, Beijing University of Posts and Telecommunications (BUPT), China. From September 2000 to June 2001, he was an Associate Professor at BUPT. From July 2001 to November 2002, he was a Research Fellow with the ARC Special Research Centre for Ultra Broadband Information Networks (CUBIN), University of Melbourne,
Australia. He is currently a professor and Dean at the School of Computer and Communication Engineering, University of Science and Technology Beijing.
He has published more than 200 papers, 20 keynote speeches, and invited talks at international and local conferences. His research interests are optical
Internet technology, new generation network technology, wireless information networks, value-added services, and secure technology of networks.
Dr. Long has been a TPC or ISC member of COIN 2003/04/05/06/07/08/09/10,IEEE IWCN2010, ICON2004/06, APOC2004/06/08, Co-Chair of the organization Committee for IWCMC2006,TPC Chair of COIN 2005/08, and TPC Co-Chair of COIN 2008/10. He was awarded by the National Science Fund for Distinguished Young Scholars of China in 2007 and selected as the Chang Jiang Scholars Program Professor of China in 2008. He is a member of the Editorial Committees of Sciences in China Series F and China Communications.
\end{IEEEbiography}

\begin{IEEEbiography}[{\includegraphics[width=1in,height=1.25in,clip,keepaspectratio]{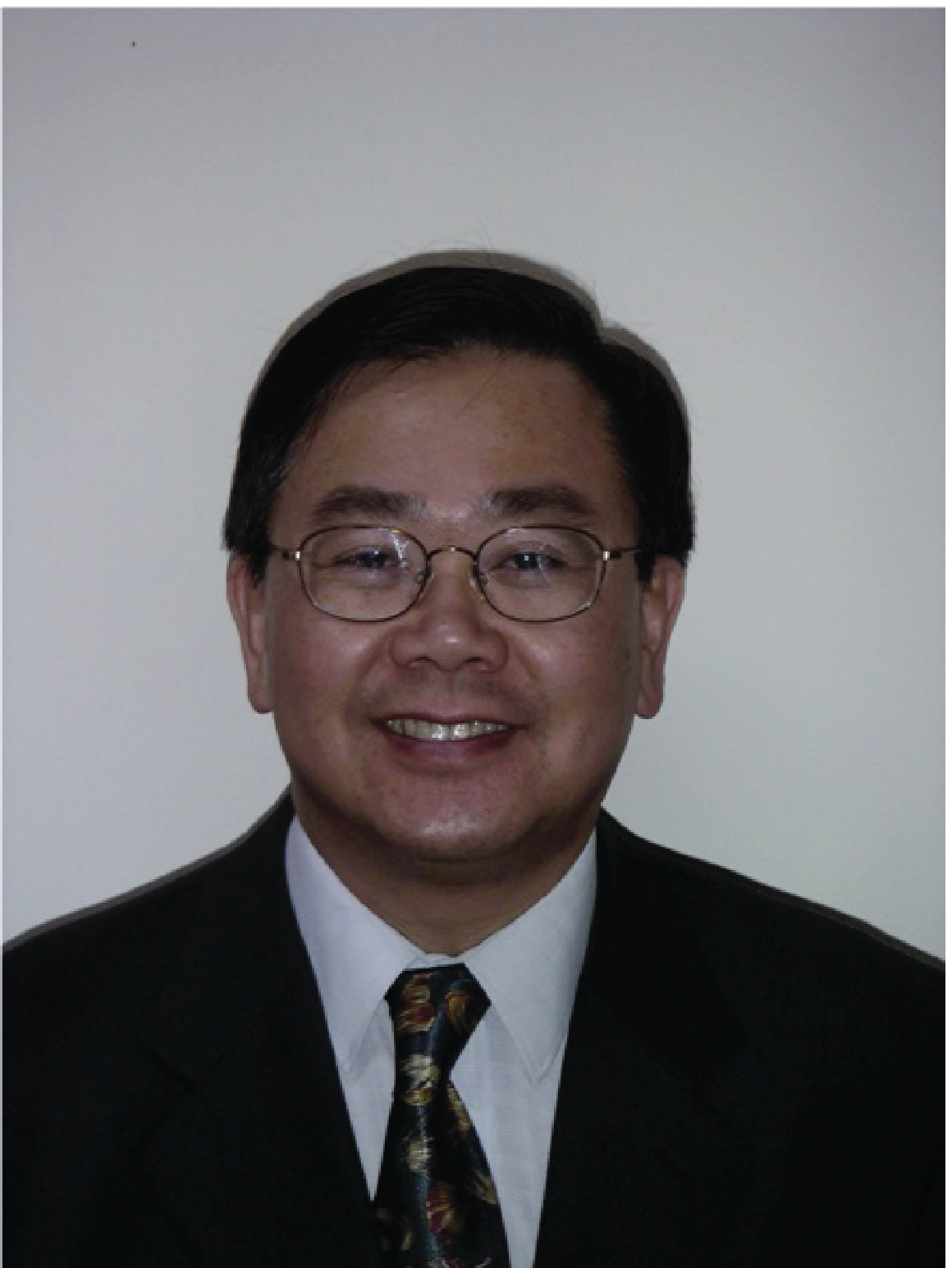}}]{Victor C. M. Leung} [S'75, M'89, SM'97, F'03] received the B.A.Sc. (Hons.) degree in electrical engineering from the University of British Columbia (UBC) in 1977, and was awarded the APEBC Gold Medal as the head of the graduating class in the Faculty of Applied Science. He attended graduate school at UBC on a Canadian Natural Sciences and Engineering Research Council Postgraduate Scholarship and received the Ph.D. degree in electrical engineering in 1982.

From 1981 to 1987, Dr. Leung was a Senior Member of Technical Staff and satellite system specialist at MPR Teltech Ltd., Canada. In 1988, he was a Lecturer in the Department of Electronics at the Chinese University of Hong Kong. He returned to UBC as a faculty member in 1989, and currently holds the positions of Professor and TELUS Mobility Research Chair in Advanced Telecommunications Engineering in the Department of Electrical and Computer Engineering. Dr. Leung has co-authored more than 1000 journal/conference papers and book chapters, and co-edited 12 book titles. Several of his papers had been selected for best paper awards. His research interests are in the areas wireless networks and mobile systems.

Dr. Leung is a registered Professional Engineer in the Province of British Columbia, Canada. He is a Fellow of IEEE, the Royal Society of Canada, the Engineering Institute of Canada, and the Canadian Academy of Engineering. He was a Distinguished Lecturer of the IEEE Communications Society. He is serving on the editorial boards of the IEEE Wireless Communications Letters, IEEE Transactions on Green Communications and Networking, IEEE Access, Computer Communications, and several other journals, and has previously served on the editorial boards of the IEEE Journal on Selected Areas in Communications - Wireless Communications Series and Series on Green Communications and Networking, IEEE Transactions on Wireless Communications, IEEE Transactions on Vehicular Technology, IEEE Transactions on Computers, and Journal of Communications and Networks. He has guest-edited many journal special issues, and provided leadership to the organizing committees and technical program committees of numerous conferences and workshops. He has received the IEEE Vancouver Section Centennial Award and 2012 UBC Killam Research Prize, and is the recipient of the 2017 Canadian Award for Telecommunications Research.
\end{IEEEbiography}
\begin{IEEEbiography}[{\includegraphics[width=1in,height=1.25in,clip,keepaspectratio]{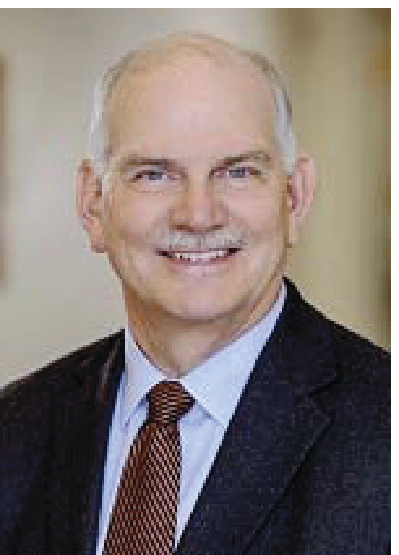}}]{H. Vincent Poor} (S'72, M'77, SM'82, F'87) received the Ph.D. degree in EECS from Princeton University in 1977.  From 1977 until 1990, he was on the faculty of the University of Illinois at Urbana-Champaign. Since 1990 he has been on the faculty at Princeton, where he is currently the Michael Henry Strater University Professor of Electrical Engineering. During 2006 to 2016, he served as Dean of Princeton's School of Engineering and Applied Science. His research interests are in the areas of information theory, statistical signal processing and stochastic analysis, and their applications in wireless networks and related fields such as smart grid and social networks. Among his publications in these areas is the book \emph{Mechanisms and Games for Dynamic Spectrum Allocation} (Cambridge University Press, 2014).

Dr. Poor is a member of the National Academy of Engineering, the National Academy of Sciences, and is a foreign member of the Royal Society. He is also a fellow of the American Academy of Arts and Sciences, the National Academy of Inventors, and other national and international academies. He received the Marconi and Armstrong Awards of the IEEE Communications Society in 2007 and 2009, respectively. Recent recognition of his work includes the 2016 John Fritz Medal, the 2017 IEEE Alexander Graham Bell Medal, Honorary Professorships at Peking University and Tsinghua University, both conferred in 2016, and a D.Sc. \emph{honoris causa} from Syracuse University awarded in 2017.
\end{IEEEbiography}
\end{document}